# Integrazione di Apache Hive con Spark

*Importazione delle collezioni da Hive in MongoDB e viceversa con l'ausilio di Hive Warehouse Connector*

Michele Gentile, Massimiliano Morrelli

Network Contacts - Via Olivetti 17 Molfetta (BA)

## Abstract

**English**. This document describes the solutions adopted, which arose from the need to transfer a large amount of information between the most famous distributed SQL and NoSQL storage systems to perform analysis and/or modification operations exploiting the peculiarities of the same.

The goal was achieved using the Spark engine and studying and using the open source library "Hive Warehouse Connector" made by Hortonworks. It provides new interoperability features between Hive and Spark.

The choice fell on these APIs in order to take advantage from Spark's distributed computing through Spark-Sql libraries, to allow a quick reading and writing on the databases chosen by the Network Contacts Systems Engineering Team and to make the stored information available for consultation outside the Ambari cluster.

Importing collections into Hive allows to query tables and perform Business Intelligence operations using the data access speed of the distributed file system.

MongoDB provides convenient access to data in the form of documents.

**Italiano.** Il presente documento descrive le soluzioni adottate, nate dalla necessità di trasferire un elevato numero di informazioni tra i più famosi sistemi distribuiti di archiviazione SQL e NoSQL per effettuare operazioni di analisi e/o modifica sfruttando le peculiarità degli stessi.

L'obiettivo è stato raggiunto utilizzando l'engine Spark e studiando e utilizzando la libreria open source "Hive Warehouse Connector" messa a disposizione da Hortonworks che fornisce nuove funzionalità di interoperabilità tra Hive e Spark.

La scelta è ricaduta su queste API per poter avvalersi del calcolo distribuito di Spark mediante le librerie di Spark-Sql, per consentire una rapida lettura e scrittura sui database scelti dal team di Ingegneria dei Sistemi di Network Contacts al fine di rendere consultabili le informazioni archiviate all'esterno del cluster Ambari.

L'importazione delle collezioni in Hive permette di effettuare query sulle tabelle ed effettuare operazioni di Business Intelligence sfruttando la velocità di accesso ai dati del file system distribuito.

MongoDB consente di accedere comodamente ai dati sottoforma di documenti.

# 1. Stato dell'arte

Durante lo studio del framework di Apache Ambari per l'analisi e la manipolazione dei big data, è stato necessario riversare il contenuto delle collezioni di MongoDB in tabelle Hive salvate su HDFS per eseguire operazioni di BI su grandi quantità di dati. Con la versione 3.0 di Hadoop i Catalog di Spark e Hive sono diventatati indipendenti. Per ovviare a questa mancanza di interoperabilità tra i due DB, Hortonworks nella seconda metà del 2018 ha rilasciato una libreria dedicata.

Le API permettono, oltre a quanto già detto, una completa integrazione delle tabelle di Hive con i Dataset di Spark. Prima dell'uscita del HWC l'accesso ai database e alle tabelle di Hive avveniva attraverso l'Hive-JDBC (Java Database Connectivity) ricalcando le modalità dei comuni database SQL.

## 1.1  Apache Ambari

Apache Ambari[1] è lo strumento open source per la gestione e il monitoraggio facilitato di un cluster Hadoop fornito direttamente da Apache. Ambari è dotato di un'interfaccia Web attraverso la quale è possibile svolgere i compiti amministrativi.

Di seguito un elenco delle componenti base integrate nell'ecosistema Apache:

- HDFS;
- YARN;
- MapReduce2;
- Tez;
- Hive;
- Pig;
- ZooKeeper;
- Ambari Infra;
- Ambari Metrics;
- SmartSense;
- Spark2;
- Zeppelin Notebook;
- Slider.

Per ognuno di questi componenti, Ambari permette di:

- utilizzare un wizard per la procedura d'installazione;
- gestire il cluster fornendo gli strumenti per far avviare e stoppare i servizi su ciascun nodo;
- effettuare il monitoraggio sia attraverso una dashboard che espone diverse metriche, sia attraverso un sistema di "alert".

> Lo strumento Ambari supporta solo versioni Hadoop 2.x e precedenti.

---

[1] La versione di Apache Ambari utilizzata è la 2.7.1.0.

Apache Ambari consente alle aziende di pianificare, installare e configurare in modo sicuro Hadoop, semplificando la manutenzione e la gestione dei cluster in corso, indipendentemente dalle dimensioni dello stesso.

## 1.2   Hive

Il software di data warehouse Apache Hive facilita la lettura, la scrittura e la gestione di set di dati di grandi dimensioni che risiedono nello storage distribuito utilizzando SQL. La struttura può essere proiettata su dati già archiviati.  Hive si integra facilmente con altre tecnologie di data center, utilizzando una familiare interfaccia JDBC.

Hive su LLAP (Live Long and Process) utilizza query persistenti con memorizzazione nella cache per evitare la latenza batch-oriented di Hadoop e fornire tempi di risposta inferiori rispetto a query che operano con un volume di dati più piccoli, Hive su Tez continua a fornire eccellenti prestazioni di query batch su set di dati su scala di petabyte.
Le tabelle in Hive sono simili alle tabelle in un database relazionale e le unità di dati sono organizzate in una tassonomia da unità più grandi a  quelle più granulari. I database sono composti da tabelle, composte a loro volta da partizioni.

All'interno di un particolare database, i dati nelle tabelle sono serializzati e ogni tabella ha una directory Hadoop Distributed File System (HDFS) corrispondente. Ogni tabella può essere suddivisa in partizioni che determinano la distribuzione dei dati all'interno delle sottodirectory della directory della tabella. I dati all'interno delle partizioni possono essere ulteriormente suddivisi in bucket.
I bucket in Hive vengono utilizzati per suddividere i dati della tabella di Hive in più file o directory rendendo più efficienti le query e hanno le segueti caratteristiche:

- I dati  presenti in quelle partizioni possono essere divisi ulteriormente in bucket;
- La divisione viene eseguita sulla base di Hash di colonne particolari che abbiamo selezionato nella tabella;
- I bucket utilizzano una qualche forma di algoritmo di hashing per leggere ogni record e posizionarlo in bucket;
- In Hive, dobbiamo abilitare i bucket usando set.hive.enforce.bucketing = true.

Hive supporta tutti i comuni formati di dati primitivi come BIGINT, BINARY, BOOLEAN, CHAR, DECIMAL, DOUBLE, FLOAT, INT, SMALLINT, STRING, TIMESTAMP e TINYINT [1].

### 1.2.1   SQL e HiveQL

Apache Hive fornisce un meccanismo per proiettare la struttura sui dati in Hadoop e per interrogare quei dati usando un linguaggio simile a SQL chiamato HiveQL (HQL).
Naturalmente ci sono un sacco di differenze tra SQL e HiveQL, ma d'altra parte ci sono anche molte somiglianze, e le recenti versioni di  Hive mantengono ancora la compatibilità SQL-92.

### 1.2.2   Hive Shell

È possibile avviare la shell di Hive, che utilizza Beeline in background, per immettere comandi HiveQL sulla riga di comando di un nodo in un cluster.
Le modifiche architetturali in Hive 3 supportano solo Beeline per l'interrogazione di Hive dalla riga di comando.
Per avviare su un cluster Hive la shell è necessario digitare *hive shell* su terminale e dare il comando di invio.
Dopo essersi connesso a Hive, viene visualizzato un prompt simile al seguente esempio:

```
Beeline version 3.1.0.3.0.1.0-187 by  Apache Hive
0: jdbc:hive2://hiveserver.hadoop.local:2181>
```

Il prompt è costituito dai seguenti componenti:

- **jdbc**: la designazione del protocollo Java Database Connectivity
- **hive2**: la designazione del protocollo HiveServer in HDP 3 per l'utilizzo di Hive 3
- **hiveserver.hadoop.local:** il nome di dominio completo (FQDN) del nodo che ospita HiveServer
- **2181**: hive.zookeeper.client.port definito in /etc/hive/conf/hive-site.xml

## 1.3   Spark

Apache Spark è una tecnologia di calcolo distribuito, progettata per eseguire elaborazioni veloci. Si basa su Hadoop MapReduce e estende questo modello per utilizzarlo in modo efficiente per più tipi di calcoli, tra cui query interattive e elaborazione di flussi. La caratteristica principale di Spark è il suo cluster computing in-memory che aumenta la velocità di elaborazione di un'applicazione.

Spark è progettato per coprire un'ampia gamma di carichi di lavoro come applicazioni batch, algoritmi iterativi, query interattive e streaming.

Il sistema Spark permette di:

- **Specificare** sequenze complete di operazioni, anche iterative, in un solo job, riducendo la complessità del codice dell'applicazione;
- **Mantenere** i dati in memoria fino al termine dell'intera sequenza di elaborazione, evitando di leggere e scrivere su disco (a meno che non fosse lo stesso sviluppatore a implementarlo);
- **Garantire** la tolleranza ai guasti dei nodi del cluster (quindi perdita di parte dei dati) assegnando l'elaborazione di un blocco di dati nello stesso nodo dove gli stessi erano già immagazzinati (*data locality*).

Apache Spark è facilmente integrabile nell'ecosistema Hadoop. Alla proprietà di *fault-tolerance* del file system distribuito si somma la velocità di elaborazione fornita dal nuovo framework.

Con Spark si è sviluppato il concetto di *Resilient Distributed Dataset* (RDD). Un *RDD* è in sostanza un set di dati suddiviso in partizioni (una tabella chiave-valore spezzata in tante sotto-tabelle o un file spezzato in tanti segmenti) che possiede alcune proprietà chiave che garantiscono il suo funzionamento:

- Ogni RDD è immutabile, cioè una volta creato non lo si può cambiare, se non creandone un altro mediante una trasformazione;
- Ogni RDD può solo essere creato inizialmente a partire dai dati su disco (presi dal HDFS) oppure a partire da altri RDD;

> Le trasformazioni possibili per creare nuovi RDD sono poche, deterministiche e ripetibili: si possono mappare (cioè trasformare un array di chiave-valore in un altro array chiave-valore), filtrare (partire da un array e crearne un altro filtrando i dati) o unire due RDD.
> Questo approccio serve perché una partizione del RDD possa essere ricostruita a partire dalla sequenza di trasformazioni che lo hanno generato.

- Ogni RDD può restare in memoria oppure essere materializzato su disco, a scelta del programmatore (RDD in memoria ma inutilizzati da tempo vengono comunque automaticamente scaricati su disco dal processo di gestione in esecuzione sul *worker*);
- Ogni RDD è descritto da un set completo di metadati che consentono la ricostruzione di una delle sue partizioni in caso di fault: dove si trovano le partizioni, quali sono gli RDD padre, quale è la sequenza di trasformazioni (*LINEAGE*) che lo hanno generato.

Spark nasce come un sistema per creare e gestire job di analisi basati su trasformazioni di RDD. Dato che gli RDD nascono e vivono in memoria, l'esecuzione di lavori iterativi o che trasformano più volte un set di dati, sono immensamente più rapide di una sequenza di MapReduce nell'ordine di 10, anche 100 volte poiché il disco non viene mai (o quasi mai) impiegato nell'elaborazione.

Spark è stato sviluppato in circa 20mila righe (rispetto a Hadoop che ne ha più di 100mila) e funziona usando HDFS o Hbase come file system distribuito e YARN o Mesos come cluster management.

Come Hadoop anche Spark si compone di un *Master Scheduler* e di tanti *worker* che eseguono il calcolo distribuito.

Il funzionamento complessivo di Spark avviene in questo modo:

1. Lo sviluppatore scrive la propria applicazione (in Java, Python, Scala...) definendo una serie di operazioni e calcoli che desidera fare sul set di dati. Nel programma possono esserci trasformazioni (generano un RDD) o azioni (fanno calcoli sul RDD e restituiscono il risultato), organizzate in sequenze o anche in cicli iterativi;

2. L'applicativo viene passato allo *Scheduler* di Spark che costruisce il grafo delle trasformazioni degli RDD;

3. In base al grafo ottenuto lo *Scheduler* determina il miglior modo possibile per distribuire i lavori di trasformazione sui nodi;

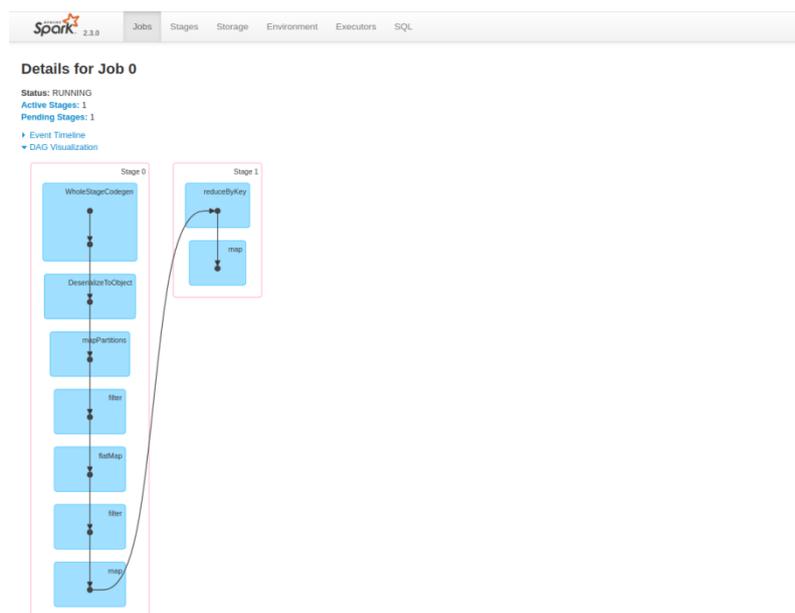

*Figura 1.1 - DAG (Directed Acyclic Graph)*

4. Al termine dell'esecuzione lo *Scheduler* ritorna il risultato finale al client che ha lanciato il programma.

### 1.3.1 Spark Shell

Spark shell è un'applicazione Spark autonoma scritta in Scala che offre un ambiente con completamento automatico in cui è possibile eseguire query ad-hoc e acquisire familiarità con le funzionalità di Spark.

È uno strumento molto utile per esplorare le diverse funzionalità disponibili in Spark con feedback immediato.

Spark Shell è uno dei tanti motivi per cui Spark è così utile per le attività di elaborazione di set di dati di qualsiasi dimensione.

Ci sono varianti di Spark shell per diverse linguaggi: spark-shell per Scala, pyspark per Python e sparkR per R [2].

Per avviare la shell di Spark è necessario digitare *spark–shell sul terminale*. Dopo essersi connesso alla shell di Spark, viene visualizzato il seguente prompt:

```
scala>
```

## 1.4    Integrazione fra Hive e Spark

A partire da HDP 3.0[2] e nelle versioni successive, Spark e Hive utilizzano catalog indipendenti per accedere alle rispettive tabelle: una tabella creata da Spark risiede nel Catalog di Spark mentre una tabella creata da Hive risiede nel Catalog di Hive e la stessa logica viene rispettata nella locazione dei database.

---

[2] L'acronimo di HDP fa riferimento a Hortonworks Data Platform

Sebbene indipendenti, queste tabelle possono interagiresolo con l'utilizzo di Hive Warehouse Connector.

*Figura 1.2 - Schema Hive WareHouse Connector*

Utilizzando Hive Warehouse Connector, è possibile esportare tabelle ed estratti dal Catalog di Spark a Hive e viceversa.
Hive Warehouse Connector supporta le seguenti applicazioni:

- Spark shell;
- PySpark;
- Spark-submit.

## 1.5 MongoDB

MongoDB è un'implementazione in C++ di un DBMS document-oriented. Sul sito ufficiale di riferimento (http://www.mongodb.org) le caratteristiche chiave del sistema vengono espresse attraverso la seguente tag-line:

> *"MongoDB (from "humongous") is a scalable, high-performance, open source, document-oriented database."*

MongoDB memorizza i dati in documenti BSON che sono molto simili ai JSON ma in rappresentazione binaria per aumentarne l'efficienza. Questo tipo di database inoltre non si basa su schemi predefiniti, infatti, le chiavi ed i valori di un documento non hanno dimensioni predefinite, il che rende la rimozione e l'aggiunta di campi molto semplice.

Nel dettaglio è un database document-oriented che sostituisce il concetto di "row" con un modello più flessibile: "il documento". Offrendo la possibilità di incorporare documenti e arrays, i db document-oriented permettono di rappresentare complesse relazioni gerarchiche con un singolo record.
I database MongoDB risiedono in un server che può ospitare uno o più database che sono indipendenti tra loro e tutti memorizzati separatamente.

Una delle caratteristiche fondamentali che distingue MongoDB dai classici db relazionali è l'approccio schemaless, ovvero l'assenza di uno schema preciso, cosa che rende la struttura di questo db molto flessibile, infatti, non è necessario come nei db relazionali definire i campi a priori e rispettare poi con rigidità le caratteristiche dei vari tipi di dato, anzi nel caso di MongoDB gli attributi possono essere modificati in maniera dinamica in un documento senza modificare le proprietà dell'intera collezione.
MongoDB ha solo una chiave obbligatoria: "_id" che corrisponde alla chiave primaria dei db relazionali e serve per identificare in maniera univoca un documento.

L'importanza delle collezioni risiede nel fatto che quando abbiamo a che fare con una grande quantità di dati, la suddivisione dei documenti in collezioni rende la ricerca molto più veloce.

### 1.5.1   La shell di MongoDB

MongoDB si basa su una shell JavaScript che permette l'iterazione con il database direttamente dalla linea di comando: eseguendo *mongo* il server del database viene attivato.

Se non viene specificato nessun comando, viene selezionato un database di default chiamato test. Per poter avere la lista dei database presenti il comando da utilizzare è *show dbs*; poi grazie al comando *use* è possibile selezionare il database sul quale lavorare.

## 2. Hive Warehouse Connector

Hive WarehouseConnector (HWC) è una libreria open source che fornisce nuove funzionalità di interoperabilità tra Hive e Spark. Hive e Spark sono spesso sfruttati dalle aziende per fornire un'infrastruttura scalabile per il data warehousing e l'analisi dei dati.

Tuttavia, poiché entrambi continuano ad espandere le loro capacità, l'interoperabilità tra i due diventa difficile.

### 2.1 Creazione di Dataframes dal set di risultati di una query llap di hive

HWC funziona come libreria assimilabile a Spark con supporto per Scala, Java e Python. Espone una API in stile JDBC agli sviluppatori Spark per l'esecuzione di query su Hive. I risultati vengono restituiti come DataFrame per qualsiasi ulteriore elaborazione / analisi all'interno di Spark. I dati vengono caricati dai daemon LLAP agli executors di Spark in parallelo, decisamente più efficiente e scalabile rispetto all'utilizzo di una connessione JDBC standard da Spark a Hive.

L'accesso da HWC è mediato da HiveServer. Il seguente diagramma mostra il flusso di dati per una query utilizzando HWC da Spark:

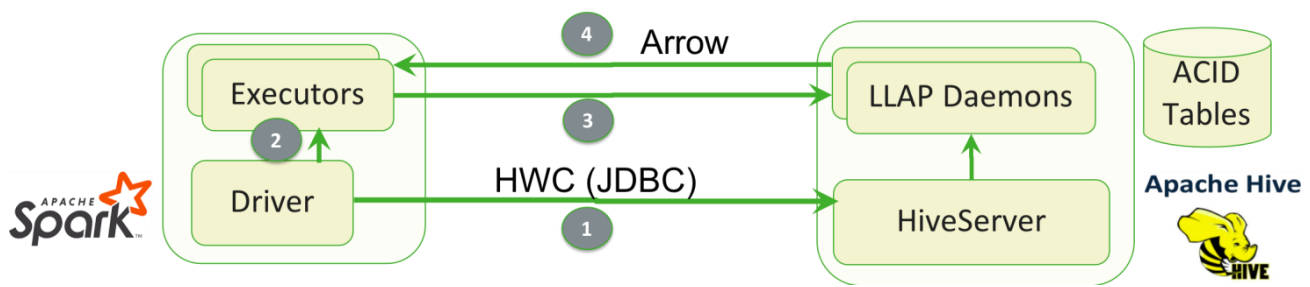

*Figura 2.1 - Flusso di dati della query utilizzando HWC da Spark*

1. Spark invia il testo sql *executequery* a Hiveserver, per ottenere un set di inputsplits da inviare a llap.
2. Gli inputsplits sono distribuiti agli executors di spark, uno split per task.
3. Ogni attività invia la propria suddivisione a un daemon llap per recuperare i dati per tale suddivisione.
4. I dati vengono scambiati tra llap e spark nel formato apache arrow. Il set di risultati è reso disponibile in modo nativo per Spark pronto per essere memorizzato nella cache, persistito o essere ulteriormente elaborato [3].

### 2.2 Scrittura di Spark Dataframes su tabelle gestite Hive

Come descritto, Spark non supporta nativamente la scrittura nelle tabelle ACID gestite da Hive. Usando HWC, possiamo scrivere qualsiasi DataFrame in una tabella del Hive Catalog. I DataFrames scritti da HWC non sono limitati per quanto riguarda la loro origine. Qualsiasi dato che è stato esposto attraverso Spark mediante l'astrazione di DataFrame può essere scritto.

HWC supporta le opzioni standard SaveMode fornite da Spark: ErrorIfExists, Append, Overwrite e Ignore.

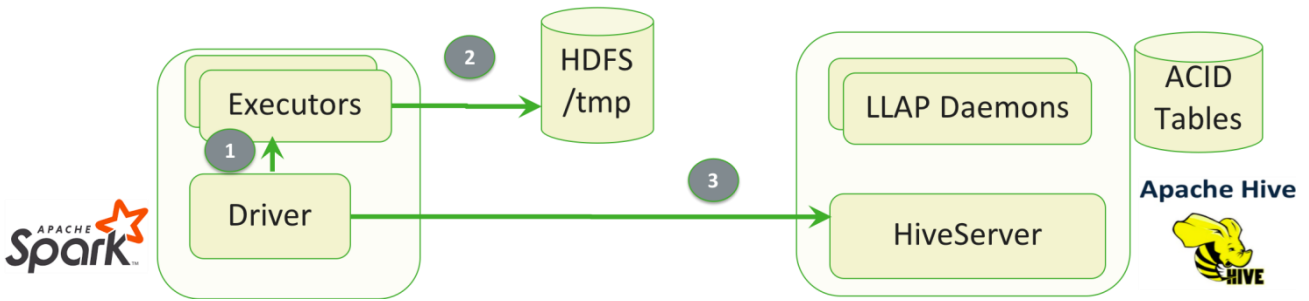

*Figura 2.2 - Flusso di scrittura di una tabella da Spark su Hive*

1. Il Driver istruirà gli esecutori di Spark che un "azione" terminale (in questo caso save ()) viene avviato su DataFrame. Ciò farà sì che Spark calcoli l'output DataFrame dalla sua linea di operazioni;
2. Il plugin HWC in esecuzione negli esecutori scriverà i risultati su storage come HDFS.
3. HWC istruirà HiveServer per caricare i dati nella tabella di destinazione [3].

## 2.3 Scritture in streaming strutturate

Spark Streaming è un'estensione di Spark, innanzitutto per eseguire un'applicazione Spark Streaming è consigliato che Spark e Kafka debbano essere distribuiti sul cluster.

Utilizzando Spark Streaming è possibile elaborare i dati utilizzando algoritmi complessi espressi con funzioni di alto livello come map, reduce, join, e window; e inviare risultati in un data warehouse, come Hive.

Spark Streaming riceve flussi di dati in input e divide i dati in batch, che vengono quindi elaborati dal motore Spark per generare il flusso finale di risultati in batch: [4]

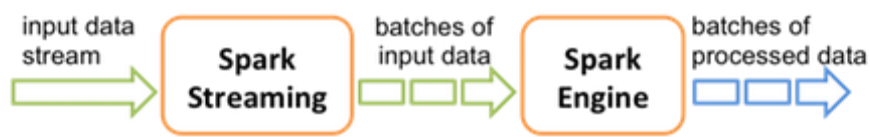

*Figura 2.3 - Flusso di dati con Spark Streaming*

HWC supporta Spark Structured Streaming nel tradizionale modello di elaborazione micro-batch.

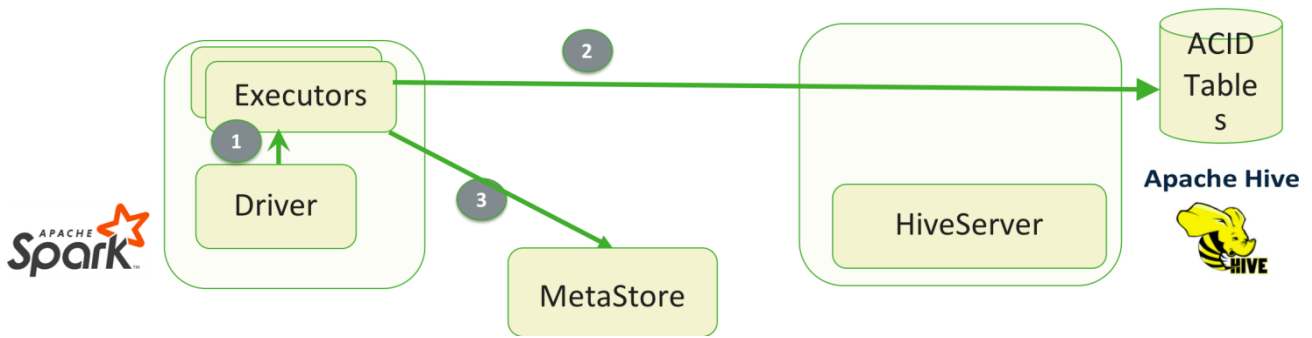

*Figura 2.4 - Scrittura in streaming utilizzando HWC*

1. Ciascun micro-batch viene inizializzato dal Driver (1);
2. I dati prodotti dai task vengono scritti in una  tabella transazionale (2);
3. Questa tabella viene committata una volta completato correttamente il batch (3).

LLAP non è richiesto per l'utilizzo del sink di streaming di HWC poiché i dati vengono scritti direttamente su disco da Spark. Il plugin di sink HWC funziona per coordinare le transazioni di streaming attraverso le operazioni di begin, commit, abort con Hive Metastore [3].

# 3. Configurazione della connessione Hive Apache <-> Spark-Apache

È possibile configurare le proprietà Spark in Ambari per utilizzare il connettore Hive Warehouse Connector permettendo un facile utilizzo di entrambi i catalog e garantendo l'interoperabilità tra i due sistemi.

## 3.1 Prerequisiti

È fortemente consigliato l'utilizzo dei seguenti software per connettere i catalog di Spark e Hive utilizzando la libreria HiveWarehouseConnector.

- HDP 3.0
- Hive con HiveServer Interactive
- Spark2

## 3.2 Proprietà richieste

È possibile configurare in modo statico le diverse proprietà necessarie alla corretta implementazione delle API HWC editando il file spark-default.conf oppure modificando la sezione *Custom spark-2-defaults* in Ambari.

In alternativa, il progettista potrà inserire o sovrascrivere la configurazione per ciascun job come illustrato nel Capitolo 5.3:

| Property | Description | Example |
|----------|-------------|---------|
| spark.sql.hive.hiveserver2.jdbc.url | URL per HiveServer2 Interactive. | jdbc:hive2:// hiveserver01.hadoop.local:2181, hiveserver02.hadoop.local:2181, hiveserver03.hadoop.local:2181;serviceDiscoveryMode=zooKeeper;zooKeeperNamespace=hiveserver2-interactive |
| spark.datasource.hive.warehouse.metastoreUri | URI per metastore. | thrift:// hiveserver01.hadoop.local:9083 |
| spark.datasource.hive.warehouse.load.staging.dir | Directory temporanea HDFS per scritture batch su Hive. | /tmp |
| spark.hadoop.hive.llap.daemon.service.hosts | Nome dell'applicazione per il servizio LLAP. | @llap0 |
| spark.hadoop.hive.zookeeper.quorum | Host Zookeeper utilizzati da LLAP. | hiveserver01.hadoop.local:2181, hiveserver02.hadoop.local:2181, hiveserver03.hadoop.local:2181 |

*Tabella 3.1- Parametri di configurazione del Job*

## 4. HiveWarehouseSession API

HiveWarehouseSession funge da API per il collegamento di Spark con Hive. Quando si implementa un job ,
è necessario istanziare un oggetto di tipo HiveWarehouseSession.

### 4.1  Implementazione in Java

Per poter utilizzare HiveWarehouseSession bisognerà importare le librerie di hortonworks e istanziare
HiveWarehouseSession.

```
import com.hortonworks.hwc.HiveWarehouseSession;
import static com.hortonworks.hwc.HiveWarehouseSession.*;
HiveWarehouseSession hive = HiveWarehouseSession.session(spark).build();
```

### 4.2  Metodi disponibili in HiveWarehouseSession

Impostare il database corrente per riferimenti tabella Hive non qualificati
```
hive.setDatabase(hiveDBName)
```

Esegue un'operazione della shell e restituisce un Dataset
```
hive.execute("describe extended hiveTableName ").show(20)
```

Mostra database
```
hive.showDatabases().show(20)
```

Mostra tabelle per il database corrente
```
hive.showTables().show(20)
```

Descrivi un tavolo
```
hive.describeTable("hiveTableName").show(100)
```

Crea un database
```
hive.createDatabase("hiveDBName",<ifNotExists>)
```

Crea una tabella ORC
```
hive.createTable("hiveTableName").ifNotExists().column("col1", "bigint"). create()
```

Cancella un database
```
hive.dropDatabase("hiveDBName",<ifExists>, <useCascade>)
```

Cancella un tabella
```
hive.dropTable( hiveTableName, <ifExists>, <usePurge>)
```

## 4.3   Operazione di lettura

Hive esegue una query di tipo SELECT e restituisce un DataSet.

```
hive.executeQuery("select * from hiveTableName ")
```

## 4.4   Operazione di scrittura

Scrivi un DataFrame per Hive in batch (usa LOAD DATA IN TABLE)

```
df.write.format(HIVE_WAREHOUSE_CONNECTOR).option("table", " hiveTableName").save()
```

Scrivi un DataFrame per Hive utilizzando HiveStreaming

```
//Using dynamic partitioning
df.write.format(DATAFRAME_TO_STREAM).option("table", hiveTableName).save()

//Or, to write to static partition
df.write.format(DATAFRAME_TO_STREAM).option("table",  hiveTableName).option("partition",
<partition>).save()
```

Scrivi uno Spark Stream per Hive usando HiveStreaming

```
stream.writeStream.format(STREAM  TO  STREAM).option("table", hiveTableName).start()
```

# 5. Integrazione Hivewarehouse Connector con MongoDB

## 5.1 Dipendenze necessarie

Per poter utilizzare i metodi messi a disposizione dalla API di HivewarehouseSession è necessario inserire nel pom.xml di un ipotetico job i riferimenti della relativa libreria. A questa è necessario affiancare le dipendenze di MongoSpark e di Spark Sql per poter usufruire rispettivamente dei metodi relativi alla lettura/scrittura su MongoDB e di quelli per la manipolazione dei dataset.  Di seguito sono riportate le dipendenze descritte.

```xml
<dependencies>
        <dependency>
                <groupId>com.hortonworks.hive</groupId>
                <artifactId>hive-warehouse-connector_2.11</artifactId>
                <version>1.0.0.3.0.1.0-187</version>
        </dependency>
        <dependency>
                <groupId>org.apache.spark</groupId>
                <artifactId>spark-sql_2.11</artifactId>
                <version>2.3.1</version>
        </dependency>
        <dependency>
                <groupId>org.mongodb.spark</groupId>
                <artifactId>mongo-spark-connector_2.11</artifactId>
                <version>2.3.1</version>
        </dependency>
</dependencies>
```

## 5.2 Repository Hortonworks

Per poter utilizzare le dipendenze di Hortonworks è necessario aggiungere il relativo repository che permette a Maven di effettuare il download delle librerie. Nel dettaglio:

```xml
<repositories>
    <repository>
        <id>public</id>
        <url>http://repo.hortonworks.com/content/repositories/releases/</url>
    </repository>
</repositories>
```

## 5.3  Istanza di HiveWareHouseSession

È necessario istanziare un oggetto di tipo HiveWareHouseSession per poter usufruire dei sui metodi durante l'esecuzione di uno Spark job. L'istanza richiesta è costruita sulla base di un oggetto di tipo SparkSession correttamente configurata con le proprietà di necessari specificate nel paragrafo 3.2.
Di seguito un esempio di istanziazione di un oggetto SparkSession utilizzando i parametri di un ipotetico cluster.

```
SparkSession sparkSession = SparkSession.builder()
                .appName("NC_HiveApp")
                .config("spark.security.credentials.hiveserver2.enabled", "false")
                .config("spark.hadoop.hive.llap.daemon.service.hosts","@llap0")
                .config("spark.hadoop.hive.zookeeper.quorum", "hiveserver01.hadoop.local:2181,
                                                    hiveserver02.hadoop.local:2181,
                                                    hiveserver03.hadoop.local:2181")
                .config("spark.datasource.hive.warehouse.load.staging.dir", "/tmp")
                .config("spark.datasource.hive.warehouse.metastoreUri", thrift://
hiveserver01.hadoop.local:9083)
                .config("spark.sql.hive.hiveserver2.jdbc.url",
"jdbc:hive2:// hiveserver01.hadoop.local:2181, hiveserver02.hadoop.local:2181,
hiveserver03.hadoop.local:2181/;serviceDiscoveryMode=zooKeeper;zooKeeperNamespace=hiveserver2-
interactive")
                .enableHiveSupport()
                .getOrCreate();
```

Il parametro *spark.security.credentials.hiveserver2.enabled* sarà settato a FALSE, ipotizzando un cluster ambari non kerberized.
Successivamente si potrà istanziare HiveWareHouseSession, specificando il database di default e l'utente da utilizzare per effettuare la connessione.

```
HiveWarehouseSession hive = HiveWarehouseSession.session(sparkSession)
                                .defaultDB( "default")
                                .userPassword("hive", "123456" )
                                .build();
```

# 6. Soluzioni implementate

## 6.1 Import data from MongoDB to Hive

L'algoritmo implementato di seguito, permette di leggere una collezione presente in un database MongoDB e di scriverla in una tabella di un database del catalog di Hive sfruttando la velocità del calcolo distribuito di Spark.

Provenendo da un DB NoSQL, i documenti di MongoDB possono presentare campi di tipo Null[3]. Questo "tipo" di dato può essere gestito tranquillamente in un Dataset, ma non in Hive.

Per ovviare a questo particolare caso il Team di Ingegneria dei Sistemi di Network Contacts ha implementato una utility che permette di gestire lo scenario descritto.

```
public void importFromMongoToHive() {
                String hiveNameTable = engine.getHiveTable();

                //Acquisisco la collezione da mongo
                Dataset<Row> documents=
sparkMongo.importDocuments(engine.getMongoURI().getCollection()).persist(StorageLevel.DISK_ONLY_2());

                // Cancella la tabella su hive se è presente
                if (engine.getDropTable()) {
                        hive.dropTable(hiveNameTable, true, true);
                }
                //Creo la tabella in "memoria"
                CreateTableBuilder newTable = hive.createTable(hiveNameTable).ifNotExists();

                //metodo per l'epurazione dello schema del dataset
                Dataset<Row> newDocument = Utils.removeNullSchema(documents);
                List<String> columnTableProperty = Utils.readSchemaDocument(newDocument);

                //Associo alla tabella in memoria le colonne con le loro proprietà, nome e tipo
                columnTableProperty.forEach(columnTable -> newTable.column(columnTable.split(";")[0],
columnTable.split(";")[1]));
                //Creo effettivamente la tabella su Hive
                newTable.create();

                //Popolo la rispettiva tabella su Hive
        newDocument.write().format(HiveWarehouseSession.HIVE_WAREHOUSE_CONNECTOR).option("table",
hiveNameTable).mode(engine.getMode()).save();
        }
```

---

[3] Ad una colonna viene assegnato il tipo Null quando durante la lettura della collezione non é presente nessun tipo di dato che permette di catalogarla e poterle associare un determinato tipo di dato.

## 6.2   Import data from Hive Catalog to MongoDB

Questo algoritmo a differenza del precedente permette di leggere una tabella presente in Hive e di scriverla in un database MongoDB.

Come già detto l'utilizzo dei metodi di Spark sono necessari per poter diminuire il tempo necessario alla scrittura delle collezioni su MongoDB utilizzando l'ambiente distribuito dello stesso.

La criticità riscontrata in questo metodo è la non chiusura del job durante l'esecuzione del metodo per salvare il dataset ottenuto dalla lettura della tabella utilizzando il metodo:

```
MongoSpark.save(hiveTable, writeConfig);
```

Anche se il dataset viene copiato sul DB di Mongo il job non riconosce la fine della scrittura dello stesso e di conseguenza non termina.

Per ovviare a questa problematica è stato necessario utilizzare il metodo Import table from Hive to Spark, leggere la tabella da spark per creare un nuovo dataset e infine salvare sul DB di Mongo il dataset appena popolato.

```
public void importFromHiveToMongo() {
            //utilizzo il catalog di Spark come db di appoggio
            importFromHiveToSpark();

            //salvo il dataset recuperato leggendo il catalog di Spark db di Mongo
            sparkMongo.saveMongoDataset(hive.session().sql("select * from " + engine.getSparkTable()),
engine.getMongoURI().getCollection());

            //Cancello la tabella creata sul catalog di Spark
            hive.session().sql("drop table " + engine.getSparkTable());
    }
```

## 6.3   Import data from MongoDB to Spark Catalog

Il metodo permette di copiare una collezione da MongoDB e salvarla in una tabella presente in Spark.
Anche in questo caso si è riscontrata la necessità di epurare la struttura del dataset letto da MongoDB dalla tipologia di dato di tipo NULL.
Al contrario però non è stato necessario creare la tabella su Spark e poi popolarla con la collezione, è bastato utilizzare il metodo saveAsTable dei dataset per il salvataggio della collezione.

```
public void importFromHiveToMongo() {
                //utilizzo il catalog di Spark come db di appoggio
                importFromHiveToSpark();

                //salvo il dataset recuperato leggendo il catalog di Spark db di Mongo
                Dataset<Row> hiveTable = hive.session().sql("select * from " + engine.getSparkTable());
                MongoSpark.save(hiveTable, writeConfig);

                //Cancello la tabella creata sul catalog di Spark
                hive.session().sql("drop table " + engine.getSparkTable());
        }
```

## 6.4   Import data from Spark Catalog to MongoDB

Il metodo permette di copiare in una tabella presente in Spark e salvarla una collezione da MongoDB.

```
public void importFromSparkToMongo() {
                Dataset<Row> sparkTable =  hive.session().sql("select * from " + engine.getSparkTable()),
                MongoSpark.save(sparkTable, writeConfig);
        }
```

## 6.5   Import table from Spark to Hive

Il metodo permette far interagire il catalog di Spark con Hive, rendendo disponibile la copia della tabella da Spark a Hive

```
public void importFromSparkToHive() {
                //Leggo la tabella dal catalog di Spark e creo un dataset
                Dataset<Row> sparkTable = hive.session().sql("select * from " + engine.getSparkTable());
                //Popolo la rispettiva tabella su Hive

        sparkTable.write().format(HiveWarehouseSession.HIVE_WAREHOUSE_CONNECTOR).option("table",
engine.getHiveTable()).mode(engine.getMode()).save();
        }
```

## 6.6 Import table from Hive to Spark

Il metodo permette far interagire il catalog di Spark con Hive, rendendo disponibile la copia della tabella da Hive a Spark.

```
public void importFromHiveToSpark() {
        //Leggo la tabella dal catalog di Hive e creo un dataset
        Dataset<Row> hiveTable = hive.executeQuery("select * from " + engine.getHiveTable());
        //Popolo la rispettiva tabella sul catalog di Spark
        hiveTable.write().option("database",
engine.getSparkDB()).mode(engine.getMode()).saveAsTable(engine.getSparkTable());
    }
```

# 7. Casi di Studio

## 7.1 Analisi delle tempistiche di risoluzioni

Il tempo di risoluzione di un job è stato il primo dato analizzato, necessario a classificare l'ordine di grandezza in tempo, di risoluzione dello stesso, le tempistiche sono state acquisite impostando il numero di nodi del parametro LLAP, inoltre si è utilizzato una collezione contenente 9.085.755 record con una struttura complessa[4].

| Soluzione | Tempo di risoluzione (min,sec) |
|---|---|
| Import data from MongoDB to Hive | 08.47 |
| Import data from Hive Catalog to MongoDB | 08:22 |
| Import data from MongoDB to Spark Catalog | 07:19 |
| Import data from Spark Catalog to MongoDB | 07:00 |
| Import table from Spark Catalog to Hive Catalog | 01:08 |
| Import table from Hive Catalog to Spark Catalog | 03:12 |

*Tabella 7.1- Tempi di risoluzione di un Job con LLAP nodes pari a 2*

I dati presenti nella Tabella 7.1- Tempi di risoluzione di un Job evidenziano una differenza sostanziale nel trasferimento della collezione fra i Catalog interni all'ambiente distribuito di Ambari e le tempistiche acquisite quando il cluster si interfaccia a un db esterno all'ambiente Ambari.
Questo risultato era atteso perché sia il catalog di Spark che quello di Hive risiedono sullo stesso file system distribuito, ovvero l'HDFS.
Inoltre dalla precedente tabella si evince una tempistica triplicata quando il trasferimento avviene tra Hive e Spark.

---

[4] La struttura della collezione risulta complessa perché sono presenti tipologia di dati di tipo array o di tipo NULL, specificità risolta utilizzando un utility creata ad hoc.

# 8. Bibliografia